\documentclass[a4paper,fleqn,usenatbib]{mnras}

\usepackage[T1]{fontenc}
\usepackage{ae,aecompl}

\usepackage{graphicx}	% Including figure files
\usepackage{amsmath}	% Advanced maths commands
\usepackage{amssymb}	% Extra maths symbols

\title[LMC cluster population]{Sizing the star cluster population of the
Large Magellanic Cloud}

\author[A.E. Piatti]{
Andr\'es E. Piatti$^{1,2}$\thanks{E-mail: andres@oac.unc.edu.ar} 
\\
$^{1}$Consejo Nacional de Investigaciones Cient\'{\i}ficas y T\'ecnicas, Av. Rivadavia 1917, 
C1033AAJ, Buenos Aires, Argentina\\
$^{2}$Observatorio Astron\'omico, Universidad Nacional de C\'ordoba, Laprida 854, 5000, 
C\'ordoba, Argentina\\
}

\date{Accepted XXX. Received YYY; in original form ZZZ}

\pubyear{2017}

\begin{document}
\label{firstpage}
\pagerange{\pageref{firstpage}--\pageref{lastpage}}
\maketitle

% Abstract of the paper
\begin{abstract}
The number of star clusters that populate the Large Magellanic Cloud 
(LMC) at deprojected distances $<$ 4 deg has been recently found to be
nearly double the known size of the system. Because of the unprecedented consequences of this outcome in our knowledge of the LMC cluster formation 
and dissolution histories,
we closely revisited such a compilation of objects and found that only
$\sim$ 35 per cent of the previously known catalogued clusters has been
included. The remaining entries are likely related to stellar overdensities
of the LMC composite star field, because there is a remarkable enhancement
of objects with assigned ages older than log($t$ yr$^{\rm -1}$) $\sim$ 9.4,
which contrasts with the existence of the LMC cluster age gap; the 
assumption of a
cluster formation rate similar to that of the LMC star field does not
help to conciliate so large amount of clusters either; and nearly
50 per cent of them come from cluster search procedures known to
produce more than 90 per cent of false detections. The lack of further
analyses to confirm the physical reality as genuine star clusters
of the identified overdensities also glooms those results. 
We support that the actual size of the LMC main body cluster population  
is close to that previously known.

\end{abstract}

% Select between one and six entries from the list of approved keywords.
% Don't make up new ones.

%%%%%%%%%%%%%%%%%%%%%%%%%%%%%%%%%%%%%%%%%%%%%%%%%%

%%%%%%%%%%%%%%%%% BODY OF PAPER %%%%%%%%%%%%%%%%%%

\section{Introduction}

Recently, \citet[][hereafter B17]{bitsakisetal2017} have reported the
detection of 4850 star clusters in the Large Magellanic Cloud (LMC),
out of which 3451 are new identifications. These surprisingly large number 
of clusters strikes our previous knowledge about the size of the LMC cluster population. 
Indeed, the most frequently used catalogue of extended objects compiled 
by \citet[][hereafter B08]{betal08} contains nearly 2580 objects distributed 
throughout the same area surveyed by B17, including actual clusters, 
associations and nebulae. On the other hand, recent searches for still unrecognised 
clusters in different regions of the LMC main body have not succeeded in 
finding large amount of new clusters, but few, if any 
\citep[e.g.,][]{p16,siteketal2016,p17a}.

The extraordinary large number of new clusters listed by B17, that by
itself surpasses the total number of known B08's clusters by far, leads to important
consequences in our understanding of the cluster disruption processes. 
For instance, \citet{baetal13} predicted a 20 per cent faster cluster dissolution
rate than the standard one based on their own complete compilation of 
ages for clusters more massive than 5000 $M_\odot$. However, bearing in mind the
age distribution of the new clusters identified by B17, such a dissolution rate might be
significantly much smaller. In addition, the observationally well-documented burst of 
cluster formation  
in both Magellanic Clouds as a consequence of the mutual tidal interaction that peaked at $\sim$
2-3 Gyr  \citep[e.g.,][and references therein]{p11a,p11b,pg13}
would appear to be blurred as well. Furthermore, the outside-in formation scenario
supported by analyses of a variety of collections of data sets 
\citep[e.g.,][]{carreraetal2011,meschin14,petal2018} is not recovered, but an inside-out
one.

Beside this, we also found some internal discrepancies in B17's results that we cannot explain.
For instance, the amount of new clusters counted from their Figure 6 - assuming
a histogram resolution of $\sim$ 1.25 pc - is roughly 1800 objects; nearly half the
total number of new clusters recognised by them. 
These reasons motived us to revisit B17's results seeking for any
clue that might shed light to conciliate their outcomes with the current knowledge
on the history of the formation and evolution of the LMC star cluster population.
We approached B17's results from four different ways as described in Sections
2 to 4, and summarise the main conclusions of our work in Section 5.

\section{Cross-correlation with the B08's catalogue}

We started by simply matching the B17's list of clusters to that of B08.
In order to do that we used the {\sc tmatch} task within IRAF\footnote{IRAF is distributed by 
the National Optical Astronomy Observatories, which is operated by the Association of 
Universities for Research in Astronomy, Inc., under contract with the National 
Science Foundation.}, which provided us with two tables that include matched and unmatched
objects, respectively. The task associates any pair of coordinates (R.A., Dec.)
in the B17's list to the one in the B08's catalogue for which the distance
between them - defined as the square root of the sum in quadrature of
Relative R.A. $\times$ cos(Dec.) and Relative Dec. - is smaller than certain tolerance value. 
The latter is represented by a circle around each point in the reference
table (B08) and is set by the user. Therefore, we firstly used a radius of 0.01 deg to 
avoid multiple matchings to B08's clusters. Then, we used the remaining unmatched
objects in both B17 and B08 lists and run {\sc tmatch} with a tolerance of 0.014 deg, 
which produced again no multiple matchings. Finally, we repeated the
procedure using a radius of 0.017 deg, without multiple cross-correlations in the
outcomes. We assumed that any object in the B17's compilation located farther than
one arcmin from any B08's cluster is an object not included in B08. This is a very
relaxed constraint, because the smallest clusters in the LMC are typically of $\sim$ 0.3 
arcmin wide in radius.

\begin{figure*}
\includegraphics[width=\textwidth]{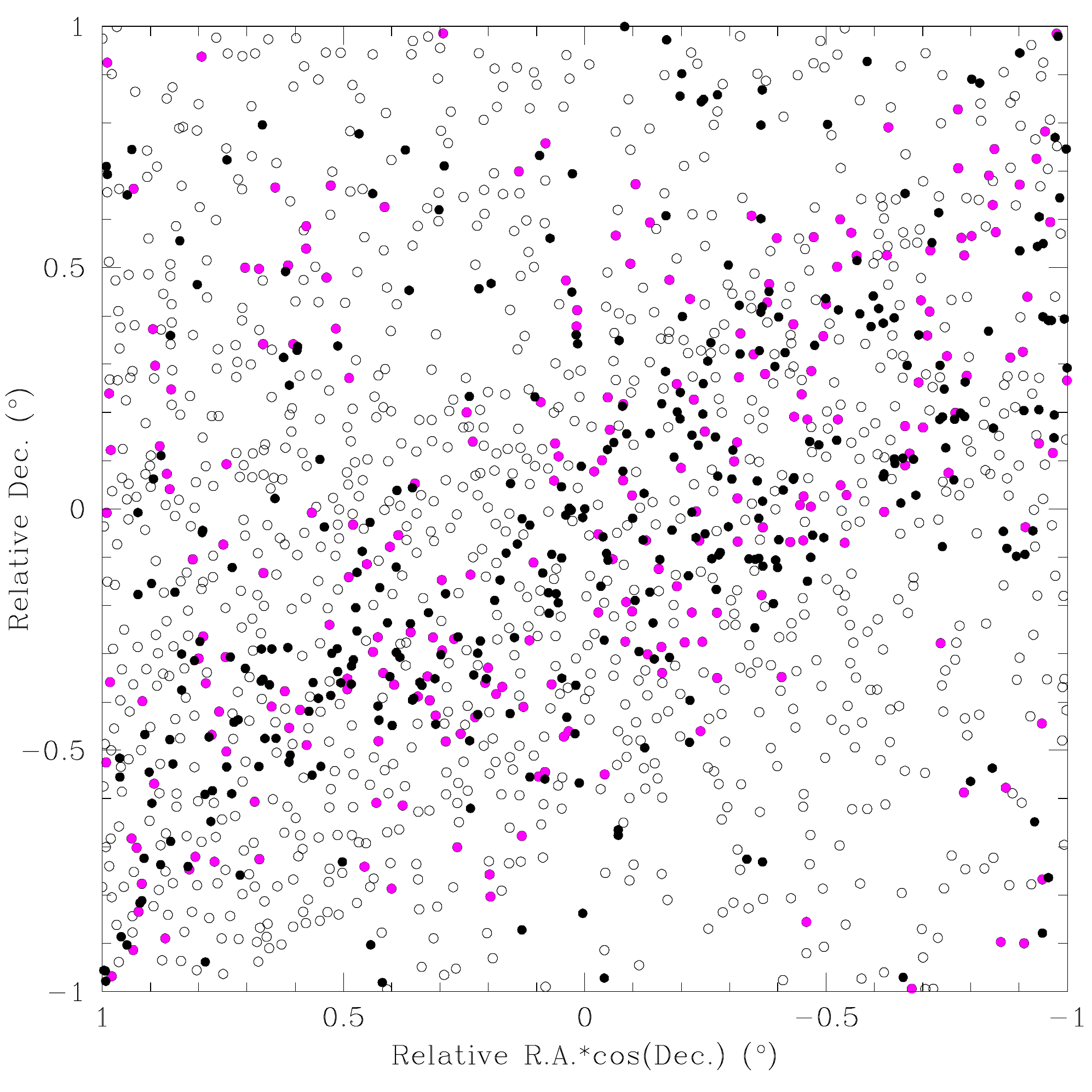}
    \caption{Schematic finding chart with the positions of matched, unmatched B08's clusters
and B17's new objects depicted with filled magenta, black and open circles, respectively, for
a central LMC bar region.}
   \label{fig:fig1}
\end{figure*}

We merged the three individual output lists with single matched objects resulting
in 918 entries. This amount of clusters represents $\sim$ 35 per cent of the
B08's clusters distributed in the area used by B17 to automatically detect resolved
objects. The remaining B17's not matched clusters add up to 3932 entries.  
Fig.~\ref{fig:fig1} illustrates the result of the matching procedure for a central
LMC area of 4 square degrees; the finding chart of the whole analysed LMC main body is
provided as on-line material for a clearer and easy inspection by the reader.
In the Appendix we show some examples of unmatched and new candidate clusters.

Two main issues arise from the above results. On the one hand, the low percentage
of recovered known star clusters and, on the other hand, the huge amount of
new detected objects, which seem somehow paradoxical. B17 did not perform
any cross-correlation with the B08's catalogue, so that they could not realise on this.
They simply run a code that identified local overdensities directly tagged as 
star clusters. Nevertheless, a stellar overdensity is far from being confirmed
as a genuine stellar aggregate without further careful analyses \citep{p14,p17e}. 
As discussed by \citet{petal2016}, the ranges of cluster dimensions and their mean 
stellar densities at different wavelength regimes play a role in automatically 
searching for star clusters, in addition to the level of photometric completeness of the 
images employed. B17 performed some Monte Carlos experiments to probe with artificial
clusters the effectiveness in recovering them, no dependence
on cluster sizes, number of stars and depth of the photometry is mentioned,
regardless the different LMC composite star field populations that contaminate
differently the cluster fields. In addition, for some of their data sets, they
yielded more than 90 per cent of false detections (see their Table 1).

\citet{pb12} showed that cleaning a colour-magnitude diagram (CMD) of a
star cluster from star field contamination is not an easy task, even more difficult 
if it is performed automatically, without any check or inspection of the resulting
cleaned cluster CMDs. The variation of the number of field stars as a function
of their magnitudes, their colour range and stochastic effects (e.g., few isolated 
bright stars) usually vary over a sky region and very frequently they 
vary around the cluster regions as well. For this reason, the general recommendation is
to use as many field areas around the cluster field as possible with relatively similar
sizes to that of the cluster area to improve the statistics \citep{metal14,maiaetal2016}.
By employing one relatively small circular star field area (see Figure 3 in B17) 
could lead to an unrepresented field population, and hence, to decontaminated cluster 
CMDs strongly featuring the composite field star population. Furthermore, the
membership probability assigned by B17 (their equation 1) depends on the position and
size of the bins overplotted on the CMDs, although the authors used fixed values
of $\Delta$(colour,magnitude) = (0.5 mag, 1.0 mag). \citet{maiaetal2010} showed that
it is necessary to apply such a procedure many times varying the size and position
of the bins, while \citet{pb12} proposed the use of variable cells to properly take into
account the actual distribution of field stars in the cluster CMD. 

\section{deep lmc fields}

\begin{figure}
\includegraphics[width=\columnwidth]{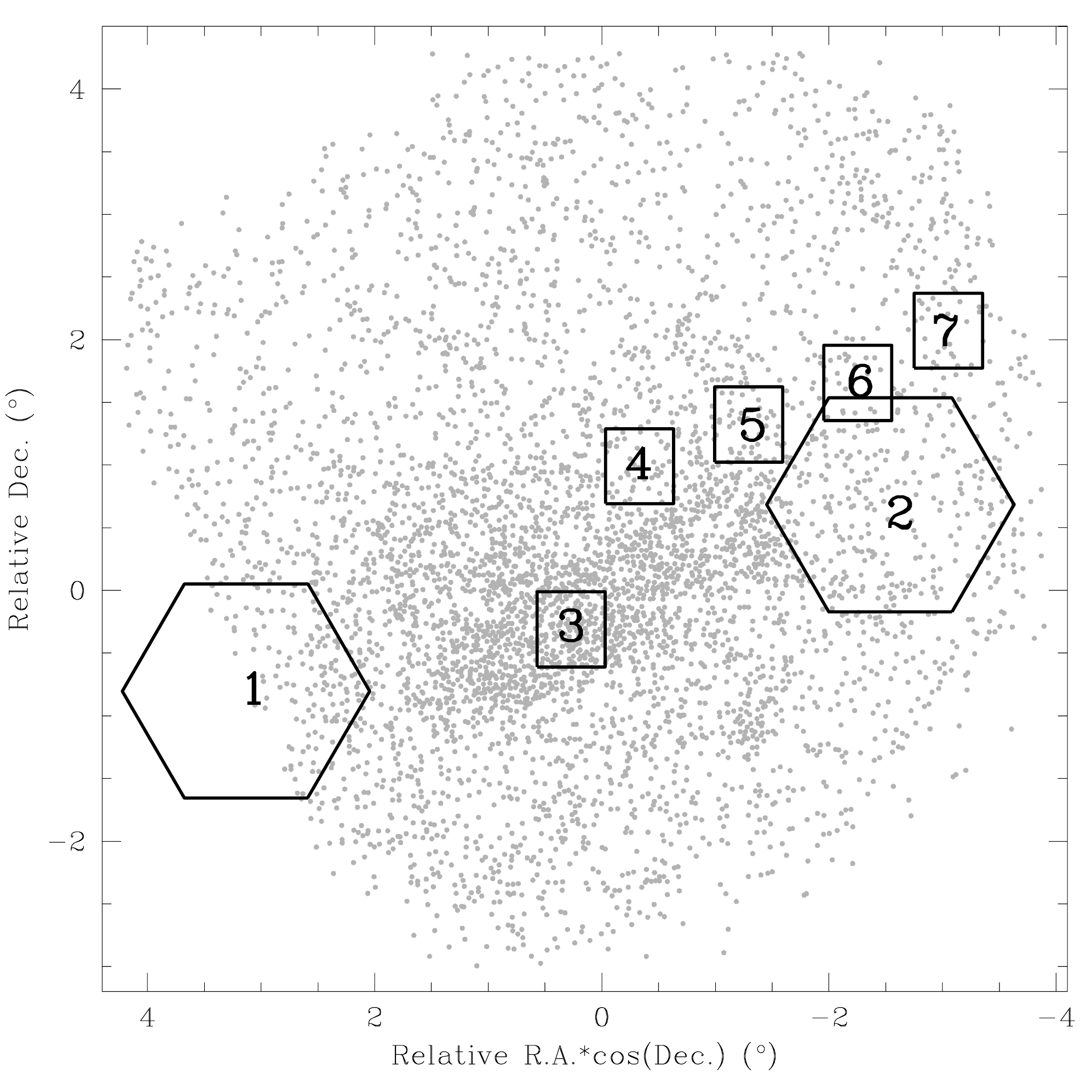}
    \caption{Spatial distribution of B17's objects with selected LMC fields studied by
\citet{p17a} (hexagons) and \citet{p17e} and \citet{petal2018} (rectangles) overplotted.}
   \label{fig:fig2}
\end{figure}

In order to closely revisit the B17's list of detected clusters, we took advantage of
recent studies that thoroughly examined the LMC cluster population in some selected
fields located within the B17's surveyed area. Fig.~\ref{fig:fig2} shows the 
distribution of the B17's objects and the seven well-studied LMC regions considered 
here overplotted. The large hexagons correspond to 
DECam fields \citep[$\sim$ 2$\degr$$\times$2$\degr$ FOV, pixel size=0.263$\arcsec$;][]{flaugheretal2015} analysed by \citet{p17a}, while the small rectangles correspond
to MOSAIC\,II fields (36$\arcmin$$\times$36$\arcmin$ FOV, pixel size= 0.269$\arcsec$)
studied by \citet{p17e} and \citet{petal2018}. The 4-m Blanco telescope at the
Cerro Tololo Interamerican Observatory (CTIO) was employed to observe those fields,
among others distributed outwards the B17's surveyed area as part of different
photometric surveys aimed at studying the most metal-poor stars outside the Milky Way 
(rectangles: CTIO 2008B-0296 programme, PI: Cole) and the Magellanic Clouds stellar history
\citep[hexagons][]{nideveretal2017}.

These LMC fields were carefully surveyed looking for star clusters by employing the 
method mentioned above \citep{petal2016}, i.e., by constraining the search using
the known cluster dimension and mean stellar density ranges. The final cluster lists
resulted to be statistically complete, because the derived photometry detected any 
star cluster based on counts of its brightest stars all the way down to
the main sequence turnoff (MSTO) of the oldest LMC clusters.  Every catalogued 
B08's cluster was recovered, in addition to very few new identified clusters in some of the
surveyed fields. The CMDs of the recognised extended objects were then cleaned from
field star contamination by using the procedure developed by \citet{pb12}, i.e., by
building statistically meaningful star field CMDs that tightly reproduced the 
luminosity function, colour distribution and stellar density of the respective
star fields, to be subtracted to the cluster CMDs. The resulting cleaned CMDs for 
some of the objects led to conclude that they are not genuine physical systems, but 
random fluctuations of the field star density. Table~\ref{tab:table1} lists the
number of entries in the B08's catalogue for each selected LMC field (field ID is
as in Fig.~\ref{fig:fig2}), and the number of true star clusters confirmed according
to the photometric analyses performed by \citet{p17a,p17e} and \citet{petal2018}.

For comparison purposes, we extracted the results of Section 2 for the selected
LMC fields, namely, the number of B08's objects matched by the B17's ones as well as
the number of B17's new detections. They are also listed in Table~\ref{tab:table1}. For 
completeness, we included in its last column the deprojected distances $d_{deproj}$ 
of the central coordinates of the seven LMC fields. They were computed by assuming 
that the LMC disc is inclined 35.14$\degr$ and has a position angle of the line 
of nodes of $\Theta$ = 129.51$\degr$ \citep{betal15}. 

From the piece of information gathered in Table~\ref{tab:table1}, we confirm previous
results by \citet{pb12} and \citet{p14}, in the sense that the B08's catalogue might 
contain $\sim$ 10-20 per cent of possible non-clusters or asterisms. Likewise, it
shows that only a couple of new genuine star clusters in some of the considered LMC
fields have been discovered, which strongly contrasts with the large number of new 
identifications compiled by B17. Because of the accuracy of the photometry used 
and the proven methods to identify clusters and to clean their CMDs employed, also 
used elsewhere \citep[e.g.,][]{p16,p16b,ivanovetal2017,p17b}, we are confident on 
the small number of new clusters found in those fields. Notice that the percentage 
of B08's clusters matched to the B17's list is nearly the same for all the fields 
($\sim$ 35$\%$), which can be considered as the effectiveness of recovering clusters 
by the automatic code of B17 at any deprojected distance.

\begin{table*}
%\centering
\caption{Statistics  of star clusters in LMC fields.}
\label{tab:table1}
\begin{tabular}{@{}cccccccc}\hline
Field ID & \multicolumn{2}{c}{B08's catalogue} & new & Ref. & \multicolumn{2}{c}{B17's objects} & $d_{deproj}$ \\
         & catalogued  &  actual               &  discoveries &      & B08's    &   new  &  (deg) \\
         &  objects    &  clusters             &              &      & objects & detections & \\\hline
1        &   75        &    75                 &  1           &  1   &  26      &  110     & 2.94 \\
2        &  144        &   144                 &  2           &  1   &  44      &  192     & 2.78 \\
3        &  107        &   69                  &  1           &  2   &  41      &  115     & 0.41 \\
4        &   26        &   21                  &  0           &  3   &   7      &   36     & 1.12 \\
5        &   25        &   17                  &  0           &  3   &   8      &   38     & 1.82 \\
6        &   24        &   24                  &  0           &  3   &   9      &   13     & 2.75 \\
7        &   11        &   10                  &  1           &  3   &   4      &   20     & 3.62 \\\hline
\end{tabular}

\noindent Ref.: 1) \citet{p17a}; 2) \citet{p17e}; 3) \citet{petal2018}.
\end{table*}

\section{star cluster frequency}

\begin{figure}
\includegraphics[width=\columnwidth]{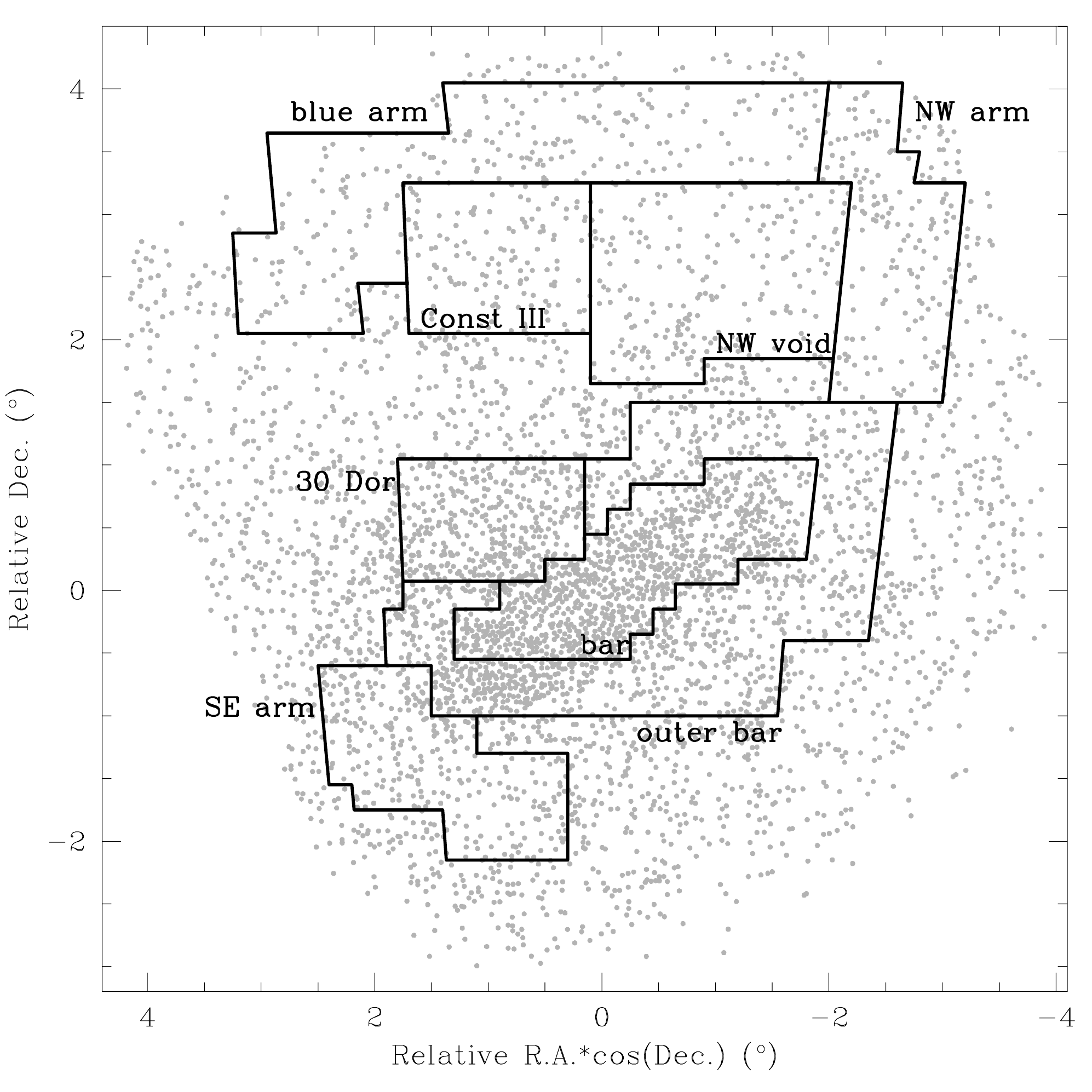}
    \caption{HZ09 regions superimposed on the B17's star cluster
spatial distribution.}
   \label{fig:fig3}
\end{figure}

With independence of drawing conclusions on the number of clusters in the
B17's compilation, it is possible to study the distribution of the
ages estimated by them in the light of the presently known LMC cluster 
formation history. In order to do that, we built the cluster frequencies
(CFs) for different regions as defined by \citet[hereafter HZ09][]{hz09}. 
Fig.~\ref{fig:fig3}
depicts those regions superposed on the spatial distribution of B17's star
 cluster. CFs are
distribution functions that trace the number of clusters per time unit
as a function of age. They are more powerful than age histograms,
because they do not depend on any age interval and the number of clusters
at different ages can be compared consistently. In addition, 
CFs can be linked to the cluster formation rate (CFR) through the expression:

\begin{equation}
CF = CFR \times \frac{\Sigma\, m^{-2}}{\Sigma\, m^{-1}} 
\end{equation}

\noindent where $m$ is the cluster mass and the sums are computed over the
LMC cluster mass range, assuming a power-law mass distribution
with a slope $\alpha$ = -2. \citet{p14b} built CFs for the HZ09 regions using 
a statistically
complete sample of clusters with accurate age estimates. We here consider them
as representative LMC CFs. After performing
a sound analysis on the effects of not considering clusters without age
estimates as well as of not including still unidentified ones, he showed that  
there exist variations of the CFs with the position in the galaxy. 

\begin{figure*}
\includegraphics[width=\textwidth]{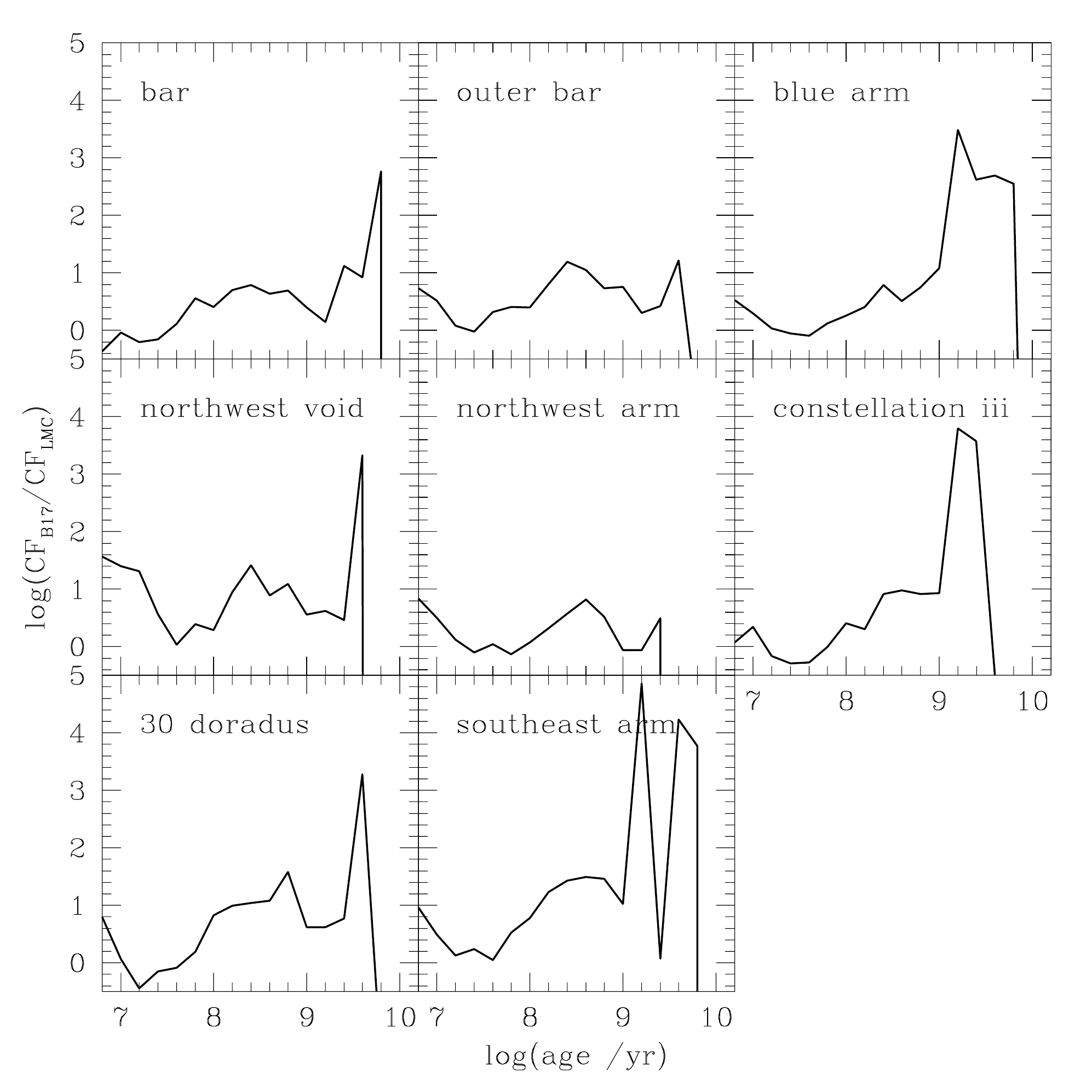}
    \caption{Relationship between the B17's CFs and those from \citet{p14b}
as a funciton of the cluster age for the different HZ09 regions.}
   \label{fig:fig4}
\end{figure*}

Following his approach, we selected the B17's clusters located in the different 
HZ09 regions and produced CFs from the number of clusters counted in age
intervals with properly chosen sizes. In doing this, we took also into account
the individual uncertainties of the age estimates.  Fig.~\ref{fig:fig4}
shows the ratio in logarithmic scale between the resulting CFs and that derived 
by \citet{p14b} as a function of age for each HZ09 region. At first glance,
noticeable differences arise from one field to another one, besides the
large excess of B17's clusters discussed in Sections 2 and 3. For instance,
the Blue Arm, Constellation III and Southeast regions present remarkable
enhancements of clusters older than log($t$ yr$^{\rm -1}$) $\sim$ 9.4,
contrarily to what would be expected, because of the well-known age gap 
($\Delta$(log($t$ yr$^{\rm -1}$)) $\sim$ 9.5 -10.0) where no LMC cluster 
apparently exist \citep[see, e.g.,][and references therein]{pg13}.
If those objects were stellar overdensities in relatively old composite star 
fields, which is a typical feature of the LMC stellar outer disc, they would
explain the excesses of objects found. Similarly, the 30 Doradus region is the one
known for its highest relative frequency of the youngest clusters, while
Fig.~\ref{fig:fig4} shows excesses of relatively old objects.

The large number of clusters older than log($t$ yr$^{\rm -1}$) $\sim$ 8.8
also contrasts with the limiting magnitudes reached by the
photometric catalogues used. \citet{zetal02} found little visible evidence 
for incompleteness for $V$ $<$ 20 mag in their Magellanic Cloud Photometric 
Survey catalogue (the deepest one used by B17), corresponding to a MSTO 
of  log($t$ yr$^{\rm -1}$) $\sim$ 8.7. Indeed, \citet{getal10}
used the same catalogue to estimate ages of 1139 LMC clusters and 
concluded that it is difficult to derive ages of star clusters older than
log($t$ yr$^{\rm -1}$) $\sim$ 9.0 due to the catalogued limited
photometric depth, which does not resolve MSTO points of 
intermediate-age and older clusters. According to the theoretical
isochrones computed by \citet{betal12}, a $\sim$ 1 Gyr old LMC cluster has its
MSTO at $V$ $\sim$ 20 mag. B17 applied an automatic technique
to fit isochrones to the observed CMDs, which could have left little space
to control stochastic effects, particularly large colour dispersion
at fainter magnitudes, photometry incompleteness at the bottom part of the
CMDs, and residuals from field star decontamination (see their Fig. 4). 
The latter could have played an important role as discussed in Section 3,
as is also the commom presence of residual LMC red clump stars that, in combination
with relatively faint main sequences, could mimic the appearance of
features typically seen in CMDs of clusters older than log($t$ yr$^{\rm -1}$) 
$\sim$ 8.7.

\section{new detections' features}

At this point, a question arises unavoidably: why did not B17 detect known 
clusters - unmatched B08's clusters - with radii and ages similar to those 
matched B08's clusters, and so many uncatalogued? We pose this question 
because in the sky,
the unmatched B08's clusters look pretty similar in average to those
matched in terms of brightness relative to the background, size, etc.
They were catalogued by visual inspection of photographic plates or 
digital images under more or less uniform detectability criteria.

\begin{figure}
\includegraphics[width=\columnwidth]{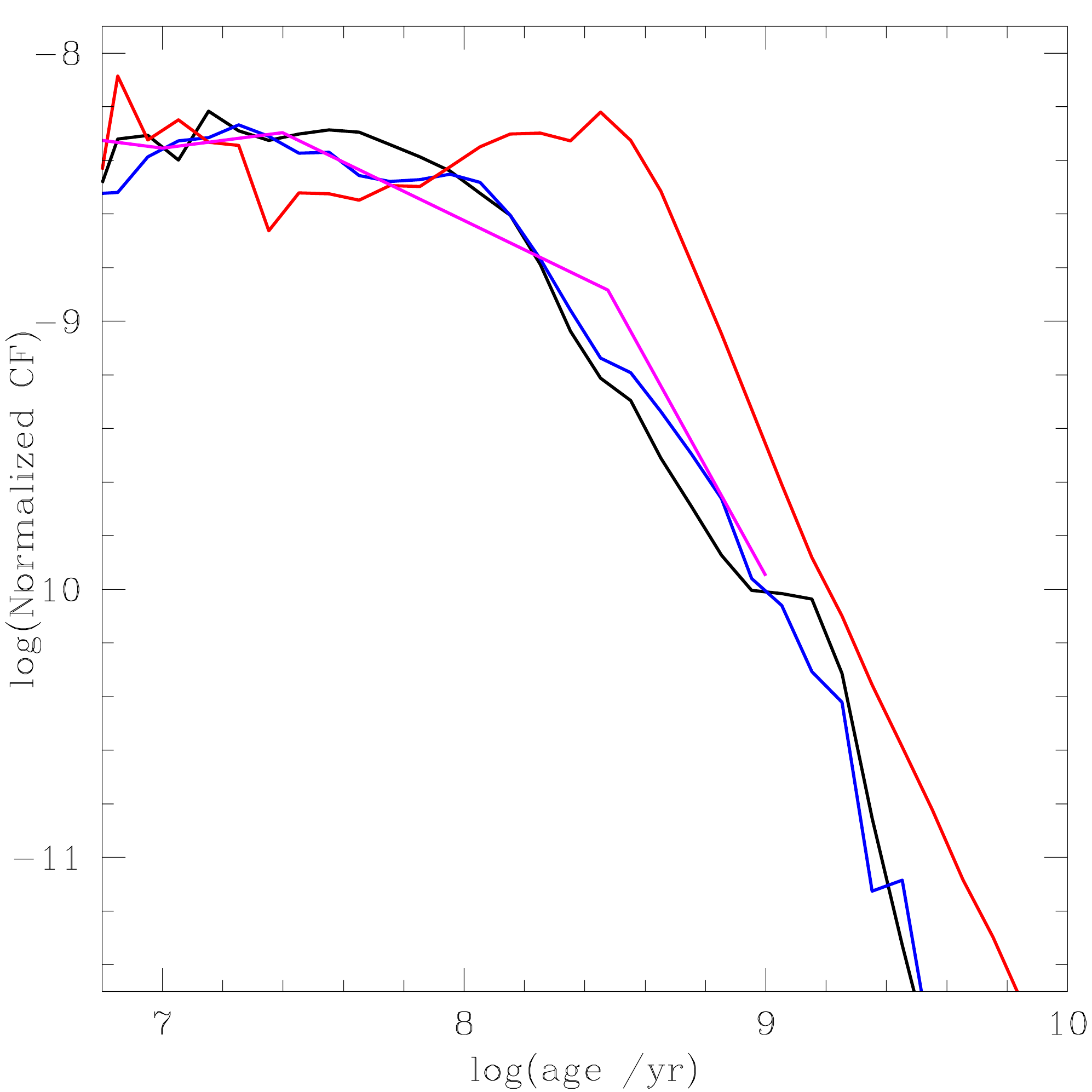}
    \caption{Normalized CFs for matched and unmatched B08's clusters
and B17 new detections represented by black, blue and red lines, respectively.
The magenta line represents the expected present-day CF if the HZ09's SFR 
and the dissolution rate derived by \citet{baetal13} are used.}
   \label{fig:fig5}
\end{figure}

In order to know whether the new identified clusters belong to a particular
kind of extended objects in the LMC, we compared the CF built from
matched B08's clusters to that from the B17's new detections. We employed
the same procedure as in Section 4, namely, considering appropriate age
bins and the age uncertainties. We divided each CF by the total number of
objects used, so that relative differences can be compared without 
introducing any shift. Fig.~\ref{fig:fig5} depicts the resulting normalized 
CFs for matched B08 and B17 new objects traced with black and red lines,
respectively. We included the normalized CF that results from using the  
unmatched B08's clusters represented by a blue line. As can be seen,
this is very similar to that coming from matched B08's objects, which means
that representative samples of the LMC cluster population (in terms of
age distribution) produce similar CFs, as expected. The resulting CF 
from the B17's new detections clearly departs from the representative
LMC CFs, with a noticeable enhancement of aged objects older than
log($t$ yr$^{\rm -1}$) $\sim$ 8.0. According to the object searching algorithm
used by B17, nearly 50 per cent  of them (1865 entries) come from detections 
made on  $Spitzer$ IRAC1 images, for which they predicted from Monte Carlo 
experiments more than 90 per cent of false detections.

The present-day CF is the result of a complex combination between the CFR and
the dissolution rate of clusters along the galaxy lifetime. If we adopted
as the CFR the star formation rate (SFR) derived by HZ09 for the LMC
stellar population, we could derive the respective CF from equation (1).
Indeed, \citet{mk2011} used only the most massive clusters on the one hand, 
and the whole cluster population on the other hand to reconstruct the LMC 
CFR, and found that there is a very good agreement between the different CFRs 
and the SFR derived by HZ09 for the last $\sim$ 1 Gyr. For older ages, the 
cluster age gap dominates the CFR, so that the SFR departs from it.
Therefore, we corrected the HZ09's SFR  from the cluster dissolution rate 
found by \citet{baetal13} - the CF/CFR ratio is 40 times smaller at
log($t$ yr$^{\rm -1}$) = 9.6 respect to that ratio at 
log($t$ yr$^{\rm -1}$) = 8.3 - to obtain the present-day CFR, and converted it 
to CF using equation (1). The normalized present-day CF for the last
$\sim$ 1 Gyr is depicted with a magenta line in Fig.~\ref{fig:fig5}. As
can be seen, there is a very good agreement with those for the
matched and unmatched B08's clusters, which suggests that the excess
of B17's new identifications older than log($t$ yr$^{\rm -1}$) $\sim$ 8.0
can possibly be related to density fluctuations of the composite
LMC stellar field.

\section{Conclusions}

We addressed the issue of the size of the LMC cluster population 
distributed throughout the main body of the galaxy ($d_{deproj}$ $<$  4 deg)
motivated by the recent compilation carried out by B17 of star clusters that 
nearly doubles the total amount of known catalogued extended objects.
Such huge amount of clusters defies our knowledge about the cluster formation
and dissolution rates, the effectiveness of past and current procedures of
identification of star clusters, etc.

We examined the B17's outcomes from four different approaches with the aim of
establishing the origin of such an enhancement of the LMC clusters
and thus reconcile them with our knowledge of the LMC cluster formation and
evolution history. 

$\bullet$ We matched the B17's list of objects
to the B08's catalogue and found that only $\sim$ 35 per cent of catalogued
extended objects in B08 have been recovered by B17. In addition, B17 identified in
average nearly 150 per cent increase of the known number of catalogued clusters.

$\bullet$ We confirmed the relative low effectiveness of the employed
cluster searching procedure by comparing the B17's findings with recent detailed
photometric studies of the actual population of genuine clusters in different
LMC regions, from the very LMC bar centre out of $\sim$ 3.6 deg. The actual
cluster populations in those selected regions were confirmed from accurate and
deep photometry that reaches the MSTO of the older LMC clusters.

$\bullet$ We built the CFs using the B17's list of objects distributed throughout
the HZ09 regions and compared them with those derived by \citet{p14b} from
a statistically complete sample of LMC clusters. Such a comparison showed that
for some of the LMC outer disc regions B17 recognized a huge amount of clusters
older than  log($t$ yr$^{\rm -1}$) $\sim$ 9.4, which contrasts with the known
cluster age gap (absence of clusters with ages in the range log($t$ yr$^{\rm -1}$) $\sim$ 9.5-10.0). Since the limiting magnitude of the deepest images used by
B17 barely reach the MSTO of clusters older than log($t$ yr$^{\rm -1}$) $\sim$
8.7, such an enhancement could be related to stellar overdensities in the
LMC composide star field.

$\bullet$ The high percentage of asterisms is also 
supported by the fact that $\sim$ 50 of them come from images for which 
Monte Carlos simulations produced more than 90 per cent of false cluster
detections. The assumption of the HZ09 SFR as CFR does not
explain such an extra amount of clusters either.

\section*{Acknowledgements}

We thank the referee for his/her thorough reading of the manuscript and
timely suggestions to improve it. 

%%%%%%%%%%%%%%%%%%%%%%%%%%%%%%%%%%%%%%%%%%%%%%%%%%

%%%%%%%%%%%%%%%%%%%% REFERENCES %%%%%%%%%%%%%%%%%%

% The best way to enter references is to use BibTeX:

\bibliographystyle{mnras}
%\bibliography{paper} % if your bibtex file is called paper.bib

%to be uncommented before sending to editor
\input{paper.bbl}

% Alternatively you could enter them by hand, like this:
% This method is tedious and prone to error if you have lots of references
%\begin{thebibliography}{99}
%\bibitem[\protect\citeauthoryear{Author}{2012}]{Author2012}
%Author A.~N., 2013, Journal of Improbable Astronomy, 1, 1
%\bibitem[\protect\citeauthoryear{Others}{2013}]{Others2013}
%Others S., 2012, Journal of Interesting Stuff, 17, 198
%\end{thebibliography}

%%%%%%%%%%%%%%%%%%%%%%%%%%%%%%%%%%%%%%%%%%%%%%%%%%

%%%%%%%%%%%%%%%%% APPENDICES %%%%%%%%%%%%%%%%%%%%%

%\newpage

\appendix

\section{Unmatched cluster and new detection examples}

\begin{figure*}
\includegraphics[width=\textwidth]{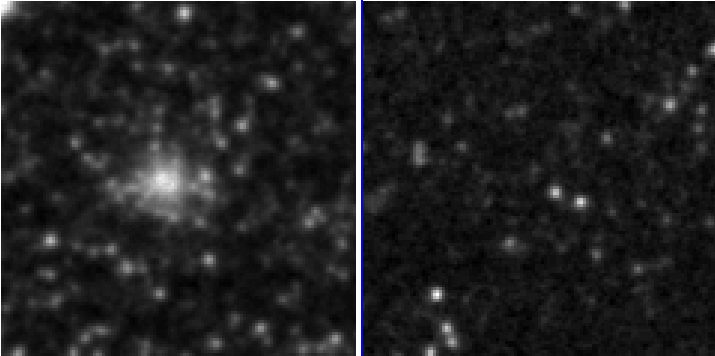}
    \caption{2$\times$2 arcmin$^2$ Red DSS images centred on the unmatched LMC globular cluster
NGC\,1928 (left) and the new detection named IR1$\_$100 (right).}
\end{figure*}

%If you want to present additional material which would interrupt the flow of the main paper,
%it can be placed in an Appendix which appears after the list of references.

%%%%%%%%%%%%%%%%%%%%%%%%%%%%%%%%%%%%%%%%%%%%%%%%%%

% Don't change these lines
\bsp	% typesetting comment
\label{lastpage}
\end{document}